\documentclass{iau} 
\usepackage{graphicx}
\usepackage[round]{natbib}

\def\DeclareAbbreviation#1#2{%
   \DeclareRobustCommand*#1{\@journalname{#2}}}
\def\@journalname#1{{\normalfont#1}}
\DeclareAbbreviation\aj{AJ}
\DeclareAbbreviation\araa{ARA\&A}
\DeclareAbbreviation\apj{ApJ}
\DeclareAbbreviation\apjl{ApJL}
\DeclareAbbreviation\apjs{ApJS}
\DeclareAbbreviation\ao{Appl.\ Opt.}
\DeclareAbbreviation\apss{Ap\&SS}
\DeclareAbbreviation\aap{A\&A}
\DeclareAbbreviation\aapr{A\&A\ Rev.}
\DeclareAbbreviation\aaps{A\&AS}
\DeclareAbbreviation\azh{AZh}
\DeclareAbbreviation\baas{BAAS}
\DeclareAbbreviation\jrasc{JRASC}
\DeclareAbbreviation\memras{MmRAS}
\DeclareAbbreviation\mnras{MNRAS}
\DeclareAbbreviation\pra{Phys.\ Rev.\ A}
\DeclareAbbreviation\prb{Phys.\ Rev.\ B}
\DeclareAbbreviation\prc{Phys.\ Rev.\ C}
\DeclareAbbreviation\prd{Phys.\ Rev.\ D}
\DeclareAbbreviation\pre{Phys.\ Rev.\ E}
\DeclareAbbreviation\prl{Phys.\ Rev.\ Lett.}
\DeclareAbbreviation\pasp{PASP}
\DeclareAbbreviation\pasj{PASJ}
\DeclareAbbreviation\qjras{QJRAS}
\DeclareAbbreviation\skytel{S\&T}
\DeclareAbbreviation\solphys{Sol.\ Phys.}
\DeclareAbbreviation\sovast{Soviet\ Ast.}
\DeclareAbbreviation\ssr{Space\ Sci.\ Rev.}
\DeclareAbbreviation\zap{ZAp}
\DeclareAbbreviation\nat{Nature}
\DeclareAbbreviation\iaucirc{IAU\ Circ.}
\DeclareAbbreviation\aplett{Astrophys.\ Lett.}
\DeclareAbbreviation\apspr{Astrophys.\ Space\ Phys.\ Res.}
\DeclareAbbreviation\bain{Bull.\ Astron.\ Inst.\ Netherlands}
\DeclareAbbreviation\fcp{Fund.\ Cosmic\ Phys.}
\DeclareAbbreviation\gca{Geochim.\ Cosmochim.\ Acta}
\DeclareAbbreviation\grl{Geophys.\ Res.\ Lett.}
\DeclareAbbreviation\jcp{J.\ Chem.\ Phys.}
\DeclareAbbreviation\jgr{J.\ Geophys.\ Res.}
\DeclareAbbreviation\jqsrt{J.\ Quant.\ Spectrosc.\ Radiat.\ Transfer}
\DeclareAbbreviation\memsai{Mem.\ Soc.\ Astron.\ Italiana}
\DeclareAbbreviation\nphysa{Nucl.\ Phys.\ A}
\DeclareAbbreviation\physrep{Phys.\ Rep.}
\DeclareAbbreviation\physscr{Phys.\ Scr.}
\DeclareAbbreviation\planss{Planet.\ Space\ Sci.}
\DeclareAbbreviation\procspie{Proc.\ SPIE}

\title[Episodic accretion in MYSOs] 
{Water masers in bowshocks: Addressing the radiation pressure problem of massive star formation}

\author[Ross A. Burns]   
{Ross A. Burns$^1$}

\affiliation{$^1$Joint Institute for VLBI ERIC (JIVE), \\ Postbus 2,
7990 AA, Dwingeloo, The Netherlands \\ email: {\tt burns@jive.nl} \\[\affilskip]}

\pubyear{2017}
\volume{336}  
\jname{Astrophysical Masers: Unlocking the Mysteries of the Universe}
\editors{A. Tarchi, M.J. Reid \& P. Castangia, eds.}
\begin{document}

\maketitle

\begin{abstract}
Ejection activities in S255IR-SMA1 and AFGL 5142 were investigated by multi-epoch VLBI observations of 22 GHz water masers, tracing bowshocks leading collimated jets. The history of ejections, revealed by the 3D maser motions and supplemented by the literature, suggests that these massive stars formed by episodic accretion, inferred via the accretion-ejection connection. This contribution centers on the role of episodic accretion in overcoming the radiation pressure problem of massive star formation - with maser VLBI and single-dish observations providing essential observational tools.  
\keywords{Masers, Stars: formation, ISM: jets and outflows}
\end{abstract}

\firstsection 
\section{Background}
 The formation of massive stars persists as one of the many exciting branches of modern astronomy. One frustration of the massive star formation community is overcoming the `radiation pressure problem' in which harsh radiation from the embedded star counteracts accretion - a problem which would limit spherical accretion models to producing stars of maximum 8 M$_{\odot}$. While non-spherical disk accretion circumvents this issue to some extent, accreting material re-encounters the radiation pressure problem at smaller radii - where gas (ionized by stellar radiation) accretes onto the star (review given in \citealt{Zinnecker07}). Such regions cannot be resolved by today's instruments. These proceedings draw on various sources from the literature to advocate episodic accretion (EA) as a means of overcoming radiation to form massive stars, and discusses observational tests of EA using VLBI observations of masers. The recipe begins with a brief introduction to its three main ingredients: the accretion-ejection relation, EA in low-mass stars, and EA in high-mass stars.

{\underline{\it [1] The accretion-ejection relation}} in young stellar objects (YSOs) is continuous over several orders of magnitude in mass and luminosity; bright accretion tracers correlate with bright ejection tracers \citep{Garatti15}. Its unbroken extension into the regime of massive star formation suggesting some degree of continuity in the physical processes governing low- and high-mass star formation, while also advocating the correlation between accretion activity and ejection activity. Via the hypothesis that each accretion event induces an ejection event, the accretion history of a young star can be inferred from its history of ejections - traced as symmetric, bipolar jet-shocks extending from the accreting object at ever increasing distances. 



\newpage

\begin{figure}[h]
\begin{center}
 \includegraphics[width=3in]{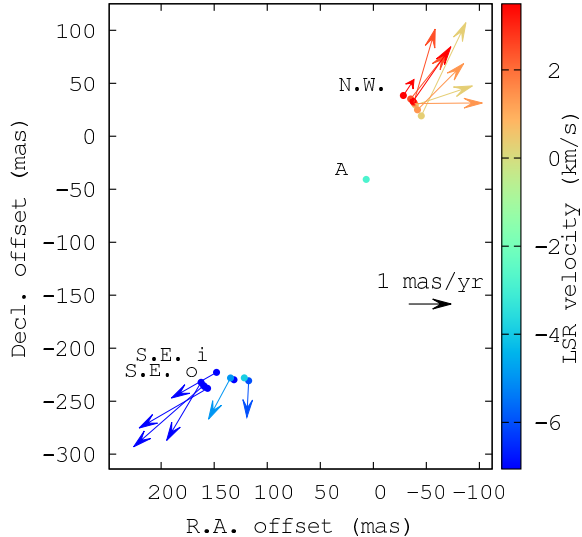} 
 \caption{Vector map of water masers in AFGL5142 MM1 from \citet{Burns17a}. North West, South East inner and South East outer bowshocks (labelled in abbreviation) extend from an MYSO located near the maser feature labelled `A'. Bowshocks indicate a junction in physical conditions generated by the most recent ejection events.}
   \label{fig1}
\end{center}
\end{figure}

{\underline{\it [2] EA in low-mass YSOs}} is exemplified observationally by the FUori and EXori classes of protostars (see review by \citealt{Audard14}). 
One mechanism by which low-mass stars might accrete episodically has been explored by \citet{Stamatellos11}, invoking a magnetic barrier which is episodically disrupted in a `magneto-rotational instability' (MRI) - leading to an accretion burst; long periods of apparent quiescence perforated by short accretion bursts, a behaviour that lends a solution to the `luminosity problem' of protostars. While single-event bursts are commonly reported in low- and high-mass stars \citep{Contreras17,Forbrich17}, establishing the episodicity of such bursts would require monitoring for long timescales; up to thousands of years. An alternative approach is therefore desirable.

{\underline{\it [3] EA in massive stars}} may provide a means of suppressing the intense radiation pressure thought to curtail accretion on to the central object. The mechanism, applicable to current-day stars but discussed in the context of primordial massive stars, is described in detail in \citet{Hosokawa16}. It is summarised as follows: Accretion of material on to the protostar causes it to `bloat'. The subsequent increase in radius leads to a drop in effective temperature. Consequently, the peak of spectral radiation (blackbody) migrates to lower frequencies, thus reducing emissivity at UV wavelengths - thereby permitting further accretion. In the absence of accretion the bloated star contracts (Kelvin-Helmholtz) toward its compact, high-temperature state, taking $\sim 10^4$ yrs. However, contraction can be stopped by further accretion events, repeatedly bloating the star and permitting further mass accumulation. This requires that accretion events occur at least every $\sim 10^4$ yrs.
Accretion events become less frequent as the reservoir of material in the envelope and disk deplete, allowing contraction to set in. UV emissivity eventually increases until accretion is finally halted and the star reaches ZAMS.

\null



Combining the topics discussed above - EA enables disk mediated accretion to form very massive stars by periodically suppressing stellar radiation (see [3]). Some (if not all) low-mass stars undergo a phase of episodic accretion (see [2]). If high-mass star formation resembles a scaled up version of low-mass star formation, as suggested by the accretion-ejection relation (see [1]), then a class of periodically accreting massive stars should be recognised - and their accretion histories can be investigated by their jet shocks.
Water masers trace such shocks, and multi-epoch observations provide their 3D motion. By combination with a parallax measurement, masers can therefore be used to accurately reveal the dynamic timescales of ejection events - and by association, accretion events - occurring in deeply embedded massive young stellar objects (MYSOs).


\vspace{-0.5cm}

\section{Observational evidence}

Episodic jets operating on timescales of $10^{3-4}$ yrs were inferred from proper motion and parallax observations of 22 GHz water masers in S255IR-SMA1 and AFGL5142. In both works VLBI water maser observations trace the youngest collimated ejections, while shock tracers at larger scales, tracing older ejections, were sourced from the literature - details given in \citet{Burns16b} and \citet{Burns17a}, respectively, with references therein. Both cases are examples of jets where the maser distributions trace clear bowshocks propagating symmetrically from the central MYSO (see Fig~\ref{fig1}); the junction in physical conditions characteristic of the onset of a new ejection.

Reports of episodic ejection in S255IR-SMA1 \citep{Burns16b} were shortly followed by an accretion burst in the same source, leading to a several magnitude increase in infrared continuum emission \citep{Garatti17}, followed by a maser burst in the radiatively pumped 6.7 GHz methanol maser line \citep{Moscadelli17}. These works demonstrate the commutation of low-mass star formation principles into the high-mass regime, namely accretion bursts (as opposed to steady accretion) and episodic jets. Furthermore, these results promote the aforementioned objects as prime candidates of MYSOs undergoing episodic accretion. Follow-up observations of masers in S255IR and AFGL5142, and other eruptive MYSOs, are underway by several groups.

\vspace{-0.5cm}

\section{Conclusions}

While EA has long been discussed in the framework of low-mass stars it has only recently been pursued in the context of massive star formation. This contribution highlights two examples of EA in MYSOs, where the inferred accretion episodes operate on periods shorter than the $10^4$ yrs required to outpace contraction, thus consistent with the mechanism described in \citet{Hosokawa16}. EA can be explored by the history of ejections from YSOs and MYSOs, providing an alternative to long-term monitoring for accretion events. Recent simulations exploring EA in massive stars were conducted by \citet{Meyer17} who also find accretion events to occur on timescales shorter than $10^4$ yr - their paper serves as a useful source of information on the topic of EA in low- and high-mass star formation. 

Previously, EA has primarily been evinced by enhancements in continuum emission. The detection of such events typically requires interferometric observations. Maser super burst events may provide an alternative approach to detecting accretion bursts and such events can be readily identified as part of maser monitoring observations conducted by single-dish radio observatories, covering large samples of sources (e.g. \citealt{Szymczak17a}). Two such maser super burst events were detected and announced during this Symposium in G25.65+1.05 and W49N (\citealt{Volvach17ATEL}, and private communication) and data from rapid follow-up VLBI observations are now being analysed. Maser observations - VLBI for investigating episodic jets, and single-dish for detecting burst events - will therefore be crucial to the integration of EA into the framework of massive star formation.






\end{document}